\documentclass[aps,prc,twocolumn,floatfix,superscriptaddress,nofootinbib]{revtex4-2}
\usepackage{lipsum}
\usepackage{graphicx}
\usepackage{verbatim}
\usepackage{multirow}
\usepackage{subfigure}
\usepackage{epsfig}
\usepackage{amsmath,amssymb,amsfonts,eufrak,mathrsfs}
\usepackage{bm}
\usepackage{color}
\usepackage{lineno}
\usepackage[bookmarks,
                bookmarksopen = true,
                bookmarksnumbered = true,
                linktocpage,
                colorlinks = true,
                linkcolor = blue,
                urlcolor  = blue,
                citecolor = blue,
                anchorcolor = green,
                hyperindex = true,
                hyperfigures]{hyperref}

\newcommand{\trento}{T\raisebox{-.5ex}{R}ENTo}
\newcommand{\RAA}{$R_{AA}$}
\newcommand{\sNN}{$\sqrt{s_{NN}}=5.02$~TeV}
\newcommand{\xsqrd}{$\chi^2_d$}

\begin{document}


\title{A simple model to investigate jet quenching and correlated errors for centrality-dependent nuclear-modification factors in relativistic heavy-ion collisions}


\author{R.\,A.\,Soltz}
\author{D.\,A.\,Hangal}
\author{A.\,Angerami}
\affiliation{Nuclear and Chemical Sciences Division, Lawrence Livermore National Laboratory, Livermore, CA 94550}

\date{\today}


\begin{abstract}

We apply Bayesian techniques to compare a simple, empirical model for jet-quenching in heavy-ion collisions to centrality-dependent jet-$R_{AA}$ measured by ATLAS for Pb+Pb collisions at \sNN.  We find that the $R_{AA}$ values for central collisions are adequately described with a model for the mean $p_T$-dependent jet energy-loss using only 2-parameters.  This model is extended by incorporating 2D initial geometry information from \trento\ and compared to centrality-dependent $R_{AA}$ values.  We find that the results are sensitive to value of the jet-quenching formation time, $\tau_f$, and that the optimal value of $\tau_f$ varies with the assumed path-length dependence of the energy-loss.  We construct a covariance error matrix for the data from the $p_T$ dependent contributions to the ATLAS systematic errors and perform Bayesian calibrations for several different assumptions for the systematic error correlations.  We show that most-probable functions and \xsqrd\ values are sensitive to assumptions made when fitting to correlated errors.
\end{abstract}

\maketitle


\section{Introduction}
\label{sec:introduction}

A wide range of models are used to study the evolution of the quark-gluon plasma created in heavy-ion collisions~\cite{10.1146/annurev-nucl-101917-020852}
at the Relativistic Heavy Ion Collider (RHIC)~\cite{Adcox:2005iw,Adams:2005gm,10.1051/epjconf/202429601006,10.1051/epjconf/202429601005,10.1051/epjconf/202429601007} at Brookhaven National Laboratory and the Large Hadron Collider at the European Center for Nuclear Research (CERN)~\cite{10.1051/epjconf/202429601001,10.1051/epjconf/202429601002}.  Over the past several years, significant progress has been made by applying Bayesian methods to multi-stage, viscous hydrodynamic models compared to a large set of single and multi-particle observables at low transverse momentum ($p_T$)~\cite{Novak:2013tf,10.1038/s41567-019-0611-8,10.1103/physrevc.103.054909,10.1103/physrevc.103.054904}, and this has yielded a common framework for exploring QGP properties in the soft (low $p_T$) physics sector.  Similar efforts are now underway to develop the analytical tools to study the properties of the QGP at high $p_T$, specifically to study the transport of high energy partons within the QGP and to understand the mechanisms by which they lose energy~\cite{10.1103/physrevc.99.054911,10.1016/j.physletb.2017.03.067,10.1007/jhep01(2020)044,10.1103/physrevc.92.011901} in a phenomena referred to as "jet-quenching"~\cite{10.1016/0370-2693(90)91409-5,10.1103/physrevlett.68.1480,10.1007/978-3-642-01539-7_17}.  The Bayesian methods that have aided our understanding of the soft sector are now being employed to the hard sector in the form of simple, few-parameter models for jet-quenching~\cite{10.1103/physrevlett.122.252302} as well as comprehensive frameworks that can be used to compare different jet-quenching physics models to a selection of jet-physics observables~\cite{10.1103/physrevc.104.024905}.  The goal of this work is to further expand upon the former, simple approach, to benefit the latter, comprehensive frameworks designed to elucidate the fundamental properties of the QGP.

There are several benefits that may be realized from this work.  The first is the ability to perform sensitivity studies to inform the inclusion and weighting of jet observables for a specific line of inquiry.  Even with significant advances in the development of computational cost-saving techniques such as Gaussian emulators~\cite{Novak:2013tf,10.1038/s41567-019-0611-8,10.1103/physrevc.103.054909,10.1103/physrevc.103.054904} and transfer learning~\cite{10.1103/physrevc.105.034910} exploratory studies may require non-negligible resources for setup and computation.  This is also true for the study of highly correlated errors where experimental guidance on the nature of the correlations may be incomplete.  Finally, access to a fast modeling platform can often stimulate new insights that can then be further tested and validated in a more sophisticated and comprehensive framework.

In this work we focus primarily on the centrality and $p_T$ dependence of the jet nuclear-modification factor, $R_{AA}$, measured for Pb+Pb collisions at \sNN, as measured by the ATLAS Collaboration using the anti-$k_T$ algorithm with jet radius, $R=0.4$~\cite{10.1103/physrevc.107.054909}.  These data were selected for the wide range in centrality and $p_T$ coverage and quantitative assessment of each of the systematic error components.  Final fit results will also be compared to similar measurements 
by the CMS and ALICE collaborations~\cite{10.1007/jhep05(2021)284,10.1103/physrevc.101.034911} collaborations.  The simple model employed is an extension of the approach adopted in~\cite{10.1103/physrevlett.122.252302}, in which the jet-energy loss distribution is described by a $p_T$ dependent function and applied to p-p jet cross-section to fit  central $R_{AA}$ values using Bayesian methods.  This work is also similar to~\cite{10.48550/arxiv.2411.14552}, which also employs a Bayesian model to simultaneously fit both \RAA\ and the di-jet assymetry, $x_J$, for central PbPb collisions to constrain jet-quenching for quark and gluon jets separately, but in this work we also extend the simple model to describe the centrality dependence of \RAA\ by incorporating geometric information provided by the \trento\  model~\cite{10.1103/physrevc.92.011901}.  The parameters are refit to the centrality-dependent \RAA\ values under two different assumptions for the correlation of systematic errors.  The jet-quenching distributions of the model are described in Sec.~\ref{sec:EDist}, and Sec.~\ref{sec:RAAcent} provides comparisons to central \RAA\ measurements.  Centrality dependent extensions and comparisons are provided in Sec.~\ref{sec:RAAcent}, and the summary and conclusions are given in Sec.~\ref{sec:summary}.
The treatment of systematic error correlations and impacts on fits are explored further in the Appendices.  In Sec.~\ref{sec:cov_compare} we compare different approaches for constructing the covariance error matrix, and the impacts of the different approaches are demonstrated in Sec.~\ref{sec:cov_impact}.

\section{Jet-quenching Distributions} 
\label{sec:EDist}

\begin{figure}[tbp]
  \begin{center}
    \includegraphics[width=0.48\textwidth]{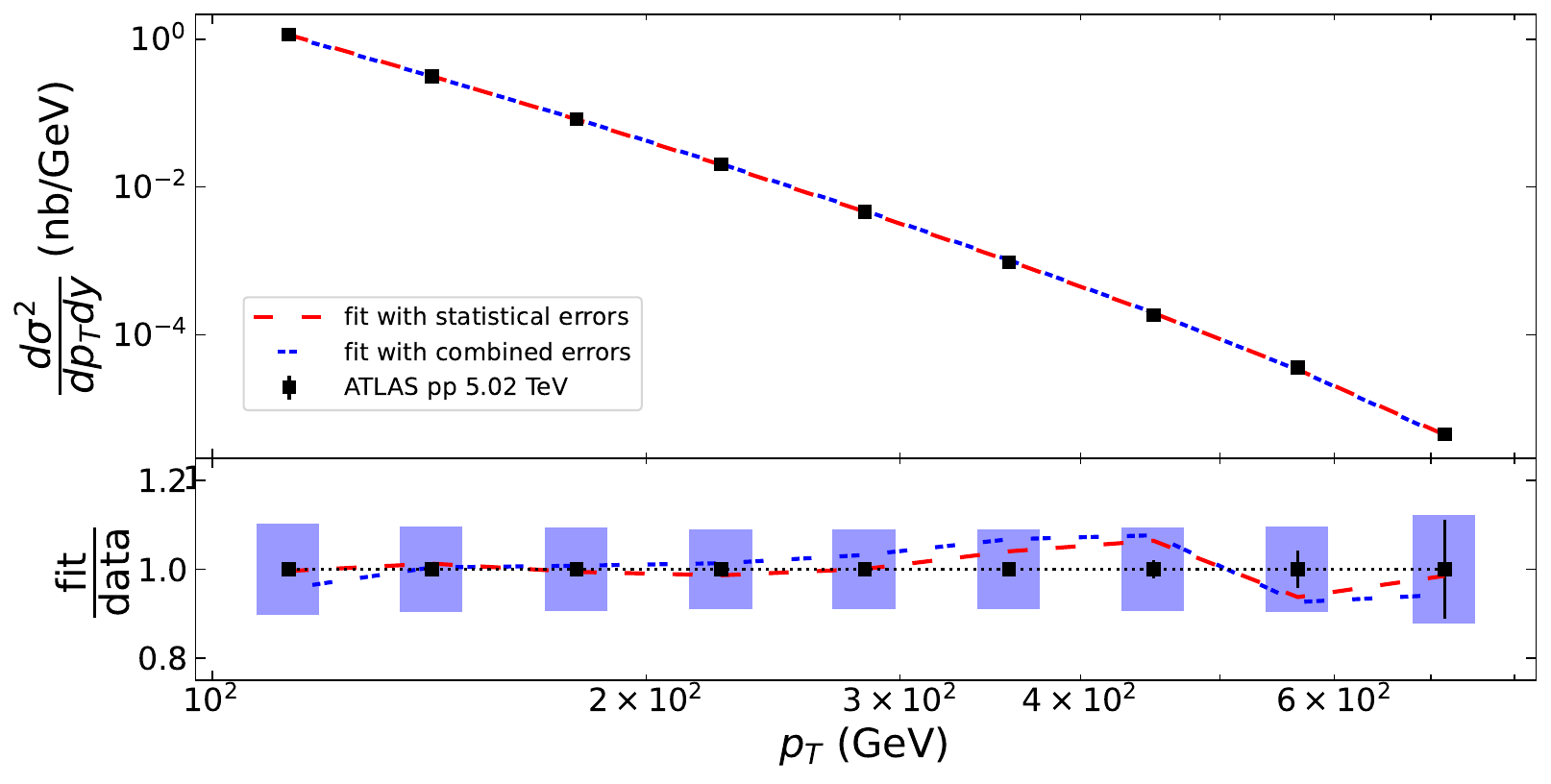}  
    \caption{ATLAS inclusive jet distributions for \sNN\ TeV p-p collisions with fits to data with statistical and combined statistical and fully-correlated systematic errors.  Ratios of fit to data shown in the bottom panel.}
        \label{fig:ppJet}
  \end{center}
\end{figure}

The first step in developing a simple model for energy loss is to parameterize unquenched, inclusive jet cross-sections in \sNN\ proton-proton collisions.  For this we use the ATLAS inclusive jet cross-section measurements in the rapidity range $|y|<2.1$~\cite{10.1016/j.physletb.2018.10.076}, and fit to Eq.~\ref{eq:ppJet} for values of $100<p_T<800$~GeV, sufficient to cover the ATLAS jet-\RAA\ measurements spanning $158<p_T<631$~GeV,
\begin{equation}
  \frac{d\sigma^2_{\rm jet}}{dp_Tdy} = \frac{(c_1/p_T)^{[c_2 + c_3 \log(p_T)]}} {1+ \exp[(p_T-c_4)/c_5]}.
\label{eq:ppJet}
\end{equation}
The terms in the numerator give a good description for $p_T$ below 500~GeV, and the denominator is used to improve agreement at higher $p_T$.  The parameters $c_4$ and $c_5$ are fixed at values of 700 and 150~GeV, respectively to improve the stability of the fits.  The fit parameters and \xsqrd\ values are given in Table~\ref{tab:ppJet} and both fit functions and data ratios are plotted in Figure~\ref{fig:ppJet}.  Although the fit with statistical errors aligns more closely to the data for nearly all $p_T$ bins, the parameterization based on the fit to the combined errors will be used in the subsequent analysis as a more faithful representation of the measurements with a proper accounting of all errors.  This choice does not impact any of the results that follow, and the topic of fitting to data with correlated systematic errors will be explored further in the remaining sections and the Appendices.
\begin{table}[tbp]
\begin{center}
   \begin{tabular}{| c | c | c | c | c |}
   \hline
   errors & $c_1$ & $c_2$ & $c_3$  & \xsqrd \\
   \hline
   statistical & 114.186(24) & 3.1261(35) & 0.50332(64) & 9.5 \\
   \hline
   combined & 103.85(24) & 2.505(30) & 0.591(79) & 3.4 \\
   \hline
   \end{tabular}
    \caption{Parameters, errors, and \xsqrd\ for Eq.~\ref{eq:ppJet} fit to \sNN\ p-p inclusive jet cross-section measured by ATLAS.}
    \label{tab:ppJet}
  \end{center}
\end{table}

The mean $p_{T}$ loss for the jet-quenching uses a form similar to the one used in~\cite{10.1103/physrevlett.122.252302}
as shown in Eq.~\ref{eq:mean_delta_pT},
\begin{equation}
  <\Delta p_T> \equiv \mu (p_T) = \alpha (p_T)^\beta \log(p_T).
\label{eq:mean_delta_pT}
\end{equation}
The distribution then takes one of two forms: a gamma-distribution with shape parameter, $k$, and scale-paramter $\theta=\mu/k$ as shown in Eq.~\ref{eq:gamma}, and a delta-distribution described by Eq.~\ref{eq:delta},
\begin{align}
f(\Delta p_T) & = \left(\frac{\mu}{k}\right)^{-k} (\Delta p_T)^{(k-1)} e^{(-\Delta p_T k)/\mu} / \Gamma(k), \label{eq:gamma} \\
f(\Delta p_T) & = \delta(p_T - \mu(p_T)) \label{eq:delta}.
\end{align}
As noted in~\cite{10.48550/arxiv.2411.14552}, the gamma-distribution is motivated by its equivalence to a convolution of $k$ exponentials, thereby providing a quasi-physical interpretation as the number of parton scatterings.  There is no physical interpretation for a delta-distribution.  It is used to test whether \RAA\ measurements are sensitive to the shape of the jet-quenching or only the mean value for a given class of events.

\section{$R_\mathrm{AA}$ for central P\lowercase{b}-P\lowercase{b}}
\label{sec:RAAcent}
\begin{figure*}[t]
  \begin{center}
    \includegraphics[width=0.80\textwidth]{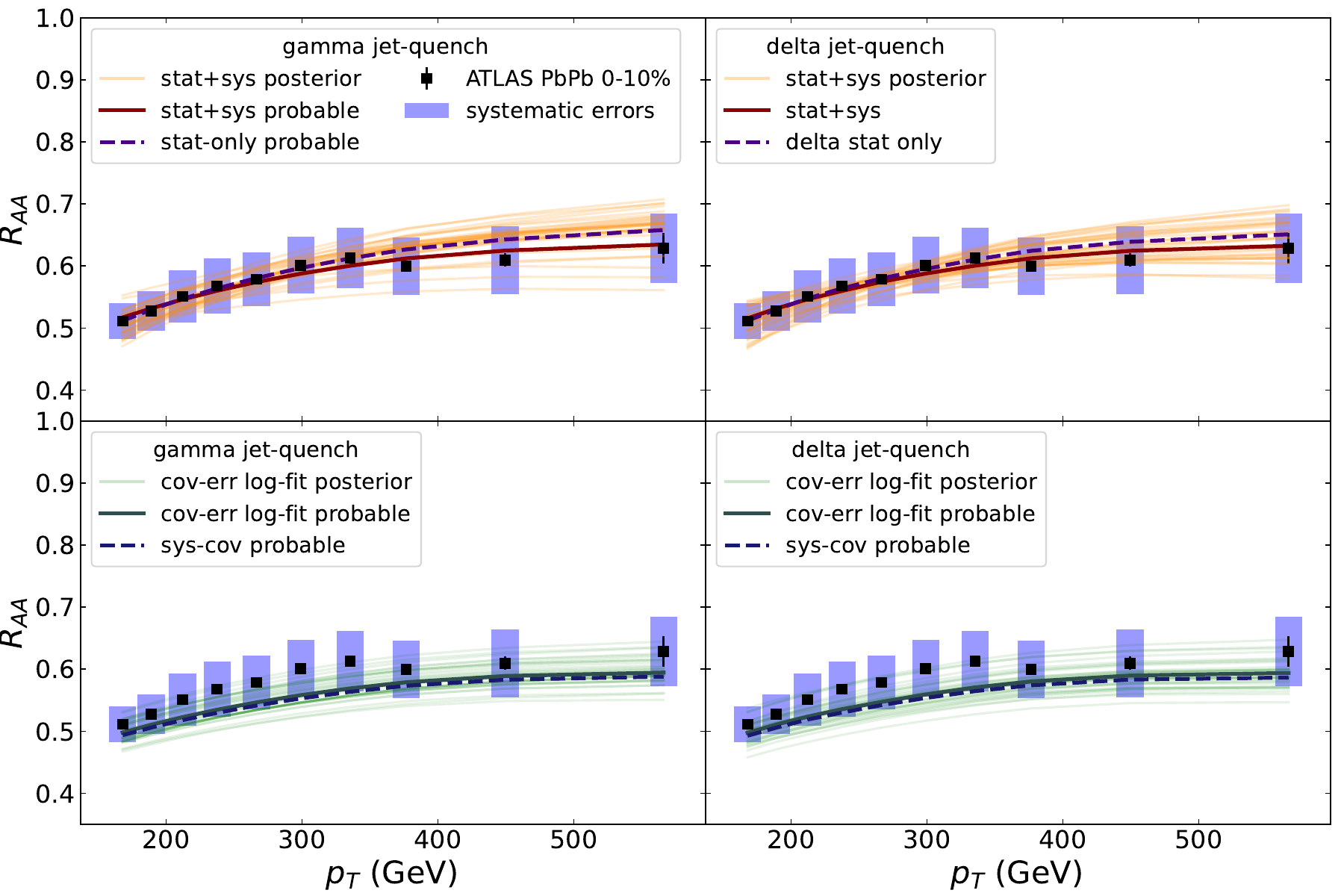}  
    \caption{ATLAS jet-RAA data, most probable fit, and 50 posterior draws for the gamma (top) and delta (bottom) functions for Bayesian fits.  The top panels show the most-probable fits using only statistical (dashed blue) and statistical plus systematic (solid red), with posterior draws for the latter.  The bottom shows the most-probable fits to full covariant error matrices (dashed blue) and fits performed taking the log of data and model (solid green) with posterior draws for the latter.} 
    \label{fig:plotRAA}
  \end{center}
\end{figure*}
Bayesian fits are performed for both the gamma-distribution and delta-distribution jet-quenching prescriptions described above.  Both formulations are fit using identical $p_T$ bins to the recently published ATLAS $R_{AA}$ measurements for 10\% central Pb-Pb collisions at \sNN\ using the emcee python package~\cite{10.1086/670067} with 10,000 samples and 8 walkers.  Parameters are drawn from uniform priors over the ranges specified in Table~\ref{tab:RAAparams}.  For each functional form, a series of four fits with different treatments of errors are shown in Figure~\ref{fig:plotRAA}.  The top panel shows the most-probable fits assuming independent errors, using only statistical (blue dashed) and statistical and systematic errors combined in quadrature in a diagonal covariance matrix (solid red).  The gamma-distribution fits are shown in the left panels and the delta-function fits are plotted on the right.  A set of 30 posterior draws for the combined errors are also shown as faint red lines in this figure.  The bottom figure shows the fits of each function to the ATLAS data with a covariance error matrix constructed by summing each of the correlated error contributions stemming from the unfolding procedure, the jet energy resolution (JER), and the jet energy scale (JES) baseline, flavor-dependent, and quenched components~\cite{10.1103/physrevc.107.054909}.  The individual contributions are each assumed to be fully correlated across all $p_T$ bins, so that the full covariance error matrix is the sum of the individual error matrices for each contribution.  The systematic error contributions from the luminosity and nuclear thickness are neglected for now and will be included in the next section on centrality dependence.   The most-probable fit using the full covariance matrix is shown as the blue-dashed line in the lower panels of Figure~\ref{fig:plotRAA}.  Fits to both the gamma and delta function prescriptions fall below the measurements, near the lower end of the systematic error bars.  This behavior is consistent with a phenomenon observed in nuclear data evaluations known as Peele's Pertinent Puzzle~\cite{10.1063/1.1945011}, in which relative-errors that are highly correlated can lead to \xsqrd\ optimizations that are systematically below the actual measurements.  One method proposed by the authors of~\cite{10.1063/1.1945011} and references therein is to take the log of the model and data before maximizing the likelihood, thereby converting relative errors into additive errors.  The solid-green lines, with 30  posterior draws are shown in the lower panels for a Bayesian calibration performed with the $\log(R_{AA}$) for both gamma (left) and delta (right) energy-loss formulations.  For both functions the most-probable $R_{AA}$ values lie slightly above the ones for which the log is not taken, but the most-probable values and most of the posterior draws still fall below the data points.  This behavior may be caused by a shape mismatch between the simple models and the measured data.  This topic is investigated further in Appendix~\ref{sec:cov_impact}.
\begin{table}[h]
\begin{center}
\begin{tabular}{| c | c | c | c | c | c |}
\hline
function & error matrix & $\alpha$ & $\beta$ & $k$ \\
\hline
\multirow{3}{*}{gamma}
& prior range & [0,10] &  [-1,10] & [1,10]  \\
& diagonal     & 0.22   & 0.18     & 7.18 \\
& covariant    & 0.14   & 0.25     & 8.19 \\
\hline
\multirow{3}{*}{delta} 
& prior range & [0,10]  & [-1,10]  &  \\
& diagonal     & 1.41    & 0.20     &   \\
& covariant    & 1.08    & 0.26     &   \\
\hline
\end{tabular}
\caption{Uniform prior ranges and maximum probability posterior values for Bayesian fits for gamma-distribution and delta-function energy loss for central jet $R_{AA}$ for \sNN\ PbPb collisions measured by ATLAS.} 
\label{tab:RAAparams}
\end{center}
\end{table}

\begin{figure}[h]
  \begin{center}
    \includegraphics[width=0.5\textwidth]{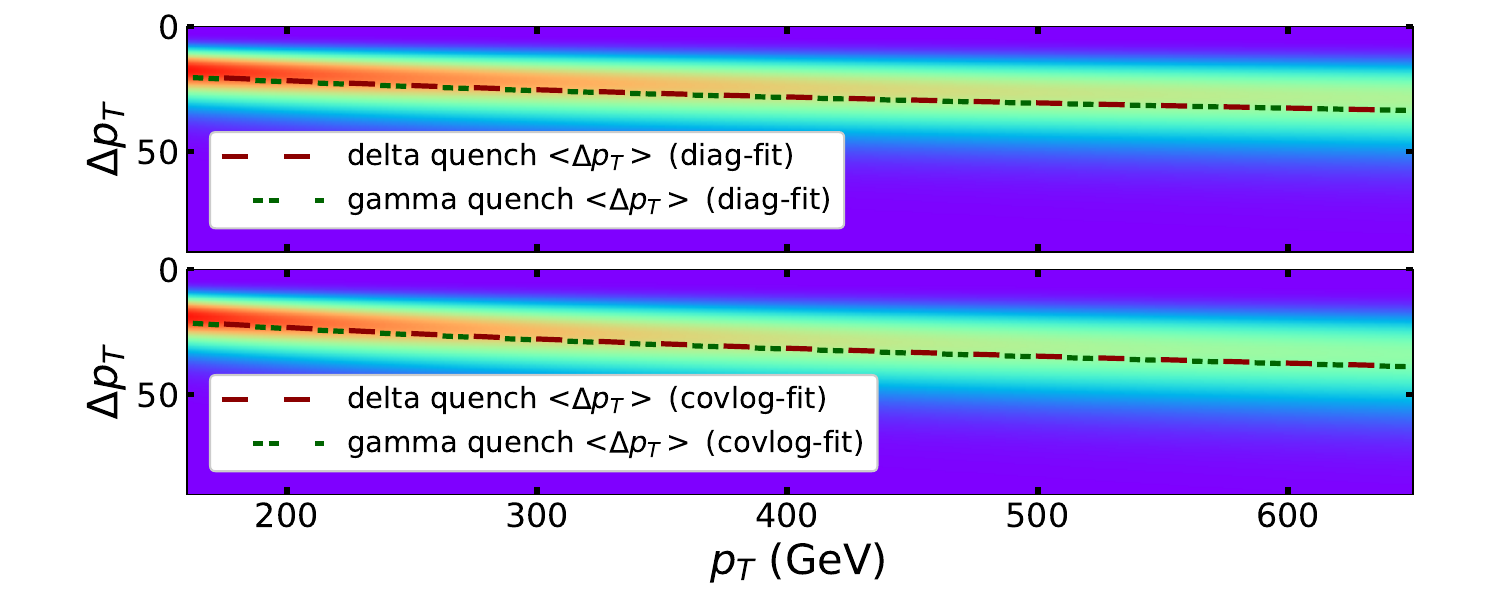}  
    \caption{Distribution for pT-loss for gamma fit shown as heatmap with most-probable mean pT-loss for gamma (dashed) and delta (dot-dashed) fits.  Upper panel shows results for fits to diagonal error matrices and lower panel uses full covariant error matrix for all $p_T$ dependent errors.} 
    \label{fig_pTquench_heatmap}
  \end{center}
\end{figure}

The mean $p_T$-loss for both functions and the gamma-distributions are shown in Figure~\ref{fig_pTquench_heatmap} for diagonal (upper panel) and full covariant errors with log-transformation (lower panel).  The mean $p_T$-loss does not vary much between the two functional forms.  The full covariant fits show large quenching above 600~GeV (nearly 40~GeV), but both fits converge to a $p_T$-loss of 20~GeV at lower $p_T$.

\begin{figure}[h]
\begin{center}
  \includegraphics[width=0.48\textwidth]{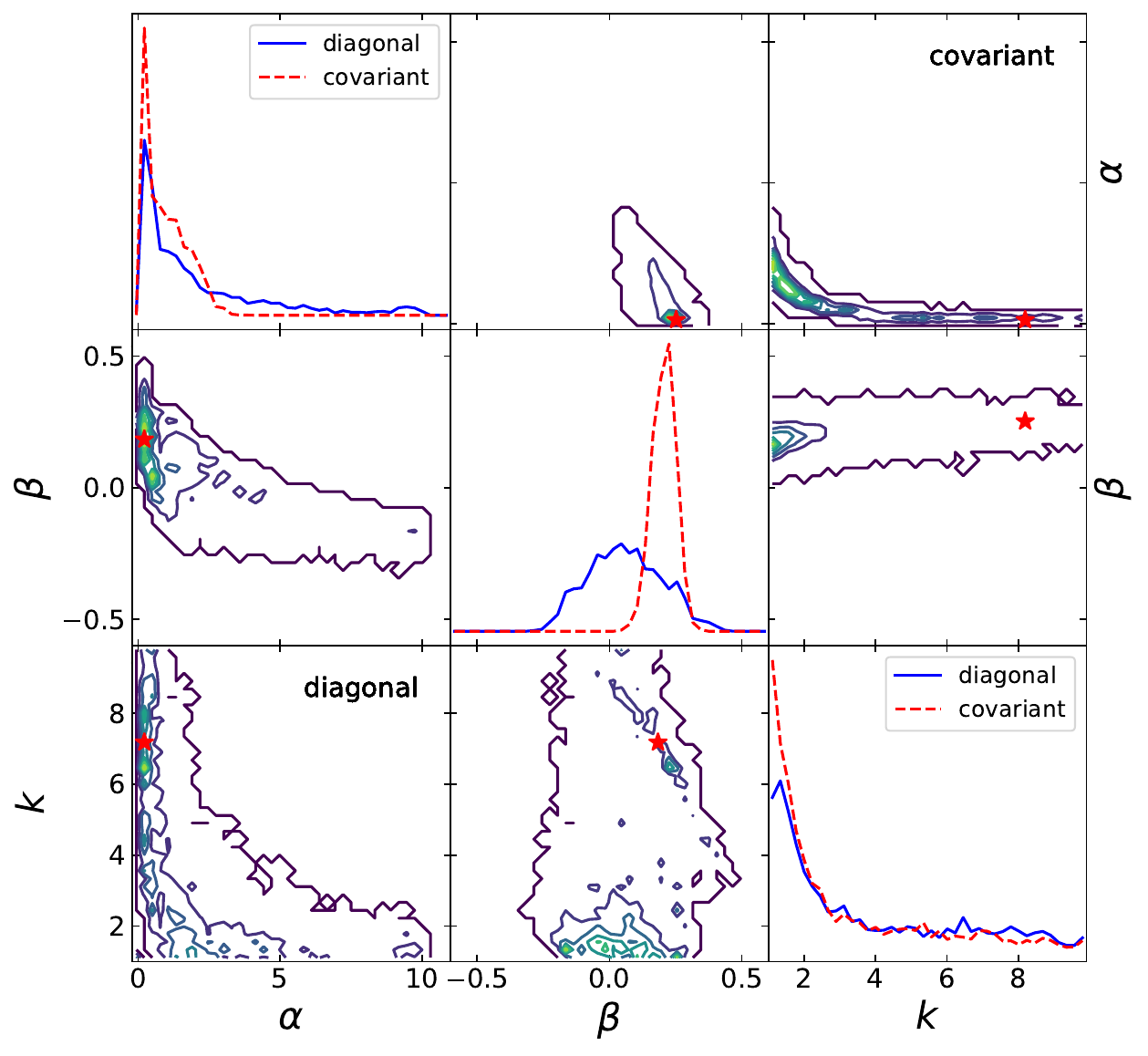}
  \caption{Corner plots for gamma-distribution fits to the ATLAS \sNN\ PbPb central jet $R_{AA}$ with diagonal errors (lower left contours and blue solid-lines) and full covariant errors    (upper right contours and red-dashed lines).  The most-probable parameter values are shown as red stars in each contour panel.}
  \label{fig:corner_gamma}
\end{center}
\end{figure}

The posterior parameter distributions and correlations for the gamma-distribution fits are shown in Figure~\ref{fig:corner_gamma}.  The parameters for the fit to combined statistical and independent systematic errors are shown in the lower left corner panels, and logarithmic fits to the data with full covariant errors are shown in the upper-right corner panels.  The single-parameter distributions are plotted along the diagonal panels with red dashed-lines for the covariant errors and solid blue lines for diagonal errors.  The $k$ parameter in the gamma-distribution is not well constrained and is strongly anti-correlated with the $\alpha$ parameter, and furthermore the most probable values, plotted as red stars, are far from the peak in the single $k$-parameter distribution.

\begin{figure}[h]
\begin{center}
\includegraphics[width=0.35\textwidth]{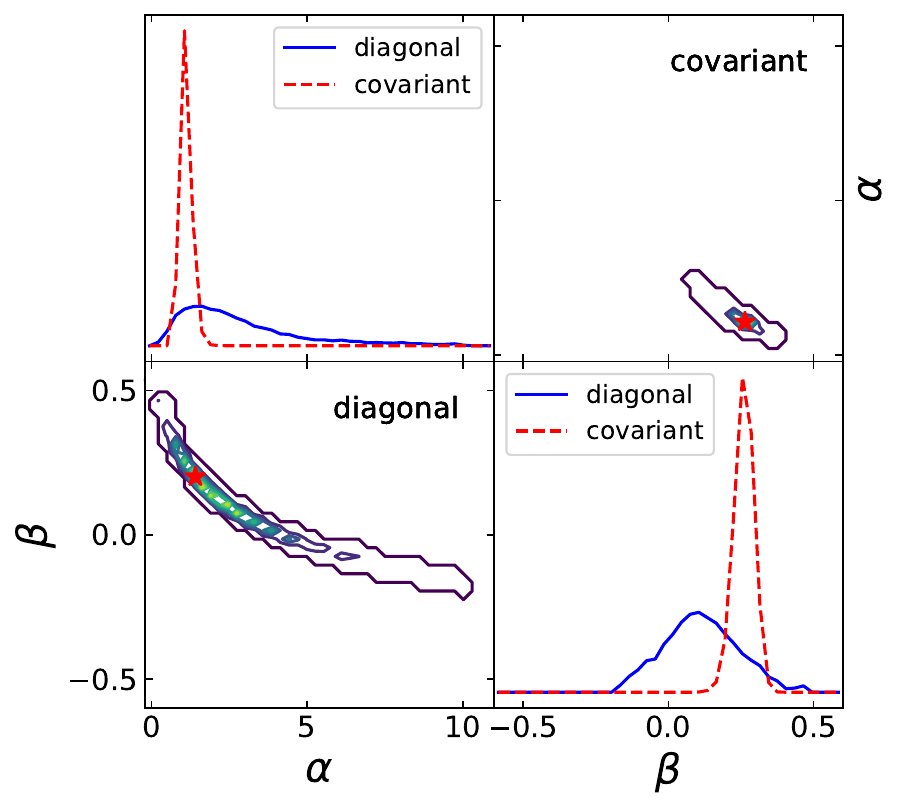}
\caption{Corner plots for delta-distribution fits to the ATLAS \sNN\ PbPb central jet $R_{AA}$ with diagonal errors (lower left contours and blue solid-lines) and full covariant errors    (upper right contours and red-dashed lines).  The most-probable parameter values are shown as red stars in each contour panel.}
\label{fig:corner_delta}
\end{center}
\end{figure}

The posterior distributions $\alpha$ and $\beta$ in delta-distribution are well-constrained, as shown in the left panels of Figure~\ref{fig:corner_delta}.  This is especially true for the fits using covariant errors plotted in the upper right corner.  Because the delta-distribution shows equivalent agreement with the measured data with better constrained parameters, we conclude that the jet-$R_{AA}$ measurements are sensitive only the the mean-value of the $p_T$-loss.   

\begin{figure}[h]
\begin{center}
  \includegraphics[width=0.35\textwidth]{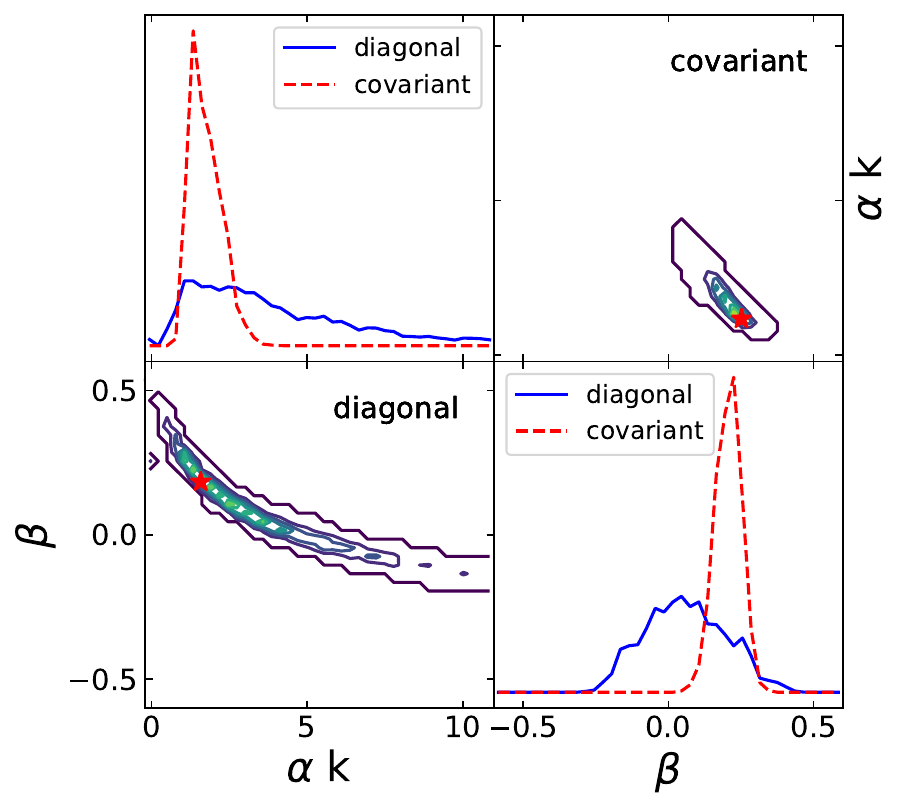}  
  \caption{Distributions and correlations for the product of the $\alpha$ and $k$ parameters for the gamma-fits to the ATLAS \sNN\ PbPb central jet $R_{AA}$ for
  with diagonal errors (blue solid-lines and upper contour) and full covariant errors (lower contour and red dashed-lines). The most-probable values are plotted as red-stars.}  
  \label{fig:corner_alphak}
\end{center}
\end{figure}

As a further comparison, the distribution for the product of the $\alpha$ and $k$ parameters is shown in Figure~\ref{fig:corner_alphak}, along with the correlation with $\beta$ for both the diagonal and covariant systematic errors.  This product is proportional to the $p_T$-independent component of the mean $p_T$-loss, and the fact that it is well-constrained is consistent with the conclusion that only the mean of the $p_T$-dependent jet-quenching is required to adequately describe measurements of jet-$R_{AA}$.  A gamma-distribution for jet-quenching can be properly constrained by also including the dijet asymmetry, $x_J$, as shown in ~\cite{10.48550/arxiv.2411.14552}.

\section{Centrality Dependent $R_{AA}$}
\label{sec:RAATrento}

\begin{figure*}[t]
  \begin{center}
    \includegraphics[width=0.9\textwidth]{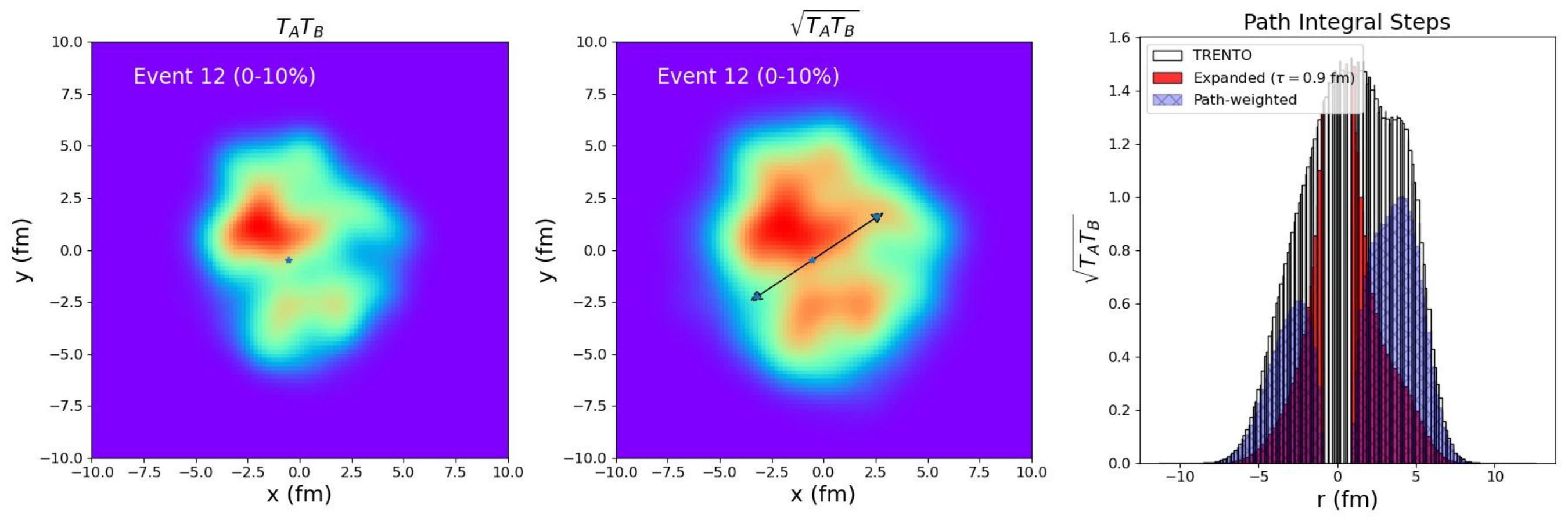}
    \includegraphics[width=0.9\textwidth]{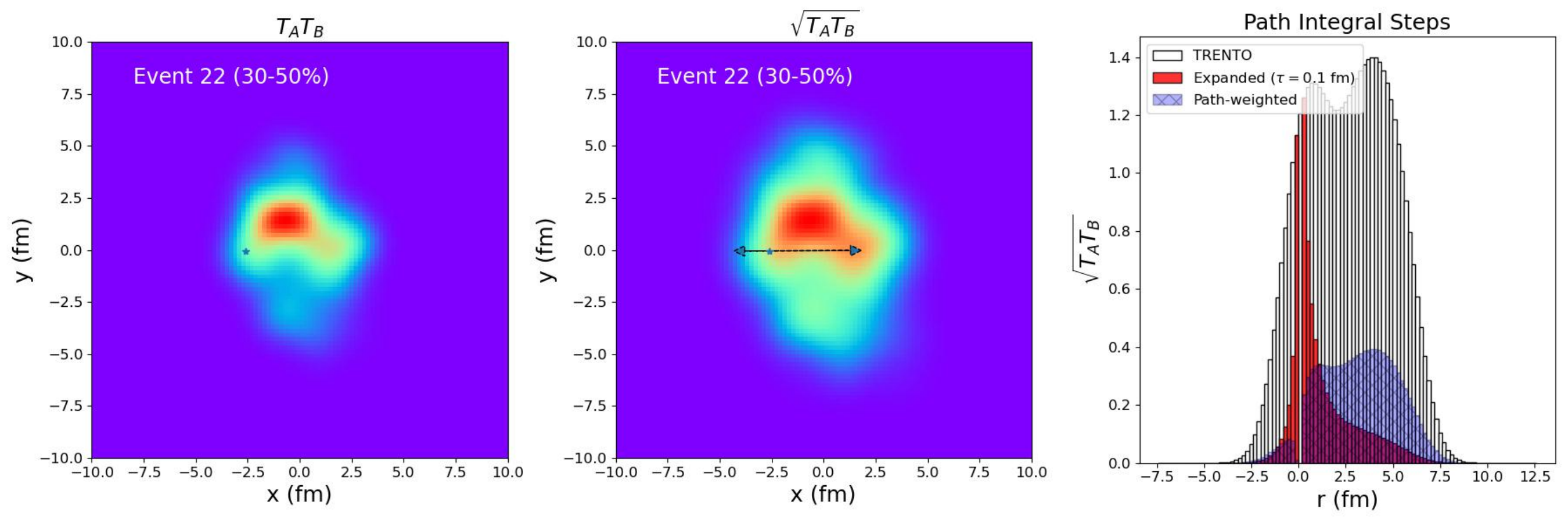}
    \caption{Distributions from \trento\ for the product of the nuclear thickness functions (left), their geometric mean (center), and the density distribution (right) along the the dijet paths shown in the central panels.  The red bars incorporate a $1/\tau=1/r$ weighting for the longitudinal expansion and the blue hatched-bars incorporate a quadratic weighting of the integrated path-lenght.  The row shows a random event in the 0-10\% centrality bin with a 0.9 fm/c formation time, and the lower row shows a random event from the 30-50\% centrality bin with a 0.1 fm/c formation time.}
    \label{fig:TATB_heatmaps}
  \end{center}
\end{figure*}

To extend the model to study the centrality dependence we use the 2-D \trento\ model~\cite{10.1103/physrevc.92.011901} to provide a simple and fast parameterization for the initial geometry in a heavy-ion collision.  The 2-D version of \trento\ generates nuclear thickness distributions in two dimensions for overlapping nuclei and uses a continuous parameter, $p$, to interpolate between the arithmetic, geometric, and harmonic means in nuclear-thickness.  These interpolations have been shown to provide good approximations to more sophisticated initial-state models~\cite{10.1103/physrevc.92.011901} and for this reason it is often used to calculate the initial state input for 2+1D multi-state hydrodynamic model comparisons to data~\cite{10.1038/s41567-019-0611-8,10.1103/physrevc.103.054909,10.1103/physrevc.103.054904}.  For this work we use parameters that are consistent with values obtained from a Bayesian fit to LHC data described in~\cite{10.1103/physrevc.103.054904}, shown in Table~\ref{tab:trento_params}.  In addition to being consistent with prior analyses, the choice of $p=0$ corresponds to the geometric mean for the nuclear thickness function, which can be squared to produce the distribution for binary nucleon-nucleon collisions that is randomly sampled to obtain the initial locations for the hard-scattered partons.  Other parameters such as total grid size (10x10~fm) and grid spacing (0.2~fm) were set to their default values.
\begin{table}[h]
\begin{center}
\begin{tabular}{| c | c | c |}
\hline
Parameter & name & value \\
\hline
$p$ & reduced thickness & 0.0 \\
\hline
$k$ & fluctuation & 1.0\\
\hline
$w$ & nucleon width        & 1.0 fm\\
\hline
$d$ & minimum distance & 1.4 fm\\
\hline
$\sigma$ & cross-section & 6.4 fm\\
\hline
\end{tabular}
\caption{\trento\ model parameters.} 
\label{tab:trento_params}
\end{center}
\end{table}

For this analysis 10,000 Pb+Pb minimum bias \trento~events were generated and separated into the ATLAS centrality bins based on participant scaling.  For each event, 10 pairs of dijet paths were generated randomly by sampling over the square of the nuclear thickness and then assigning a random azimuthal angle.  Examples of two event-paths are shown in Fig.~\ref{fig:TATB_heatmaps}, one from the 0-10\% centrality bin (upper panels) and a second event from the 30-50\% (lower panels) centrality bin.  The panels on the left show the distributions for the product of the nuclear thickness functions, from which the initial dijet locations are sampled.  The middle panels show the distribution of the \trento\ energy density to be traversed by the dijets, denoted by arrows.  The panels on the right show the energy density distributions along the two arrows as a function of the distance from the initial scattering, with the sign of the distance defined by the vertical jet-momentum component unless zero, in which case the horizontal component is used.  The energy density distributions along the dijet paths are shown as empty bars.  The energy density is then reduced according to the boost-invariant longitudinal expansion by weighting with the inverse of proper time, $\tau$, beginning after a sudden formation time, $\tau_f$, to denote the beginning of the jet-quenching.  All dijets are assumed to travel in a transverse direction with velocity $c$.  This simple approximation leads to a weighting of $\tau_f/\tau$ as illustrated by the solid red bars.  A recent study using the MARTINI model~\cite{10.48550/arxiv.2407.19966} has shown that the formation time for both the quark-gluon plasma and the jet may form at different times, but this consideration is beyond the scope of a simple model.  Because jet-quenching is expected to depend quadratically on path-length through the medium, an additional weighting proportional to $\tau-\tau_f$ is applied and plotted as hatched blue bars.

The geometry dependence of energy-loss is accounted for by weighting the delta-distribution energy loss in Eq.~\ref{eq:delta} by the path-integral sum of the expanded medium as denoted by the red bars for a linear scaling and hatched-blue bars for quadratic.  The Bayesian calibration for the $\alpha$ and $\beta$ parameters in Eq.~\ref{eq:delta} is calculated for all ATLAS centrality bins using covariance error matrices that now include the systematic error contributions for the luminosity and nuclear thickness functions.  The formation time value and choice of path-length scaling (linear vs. quadratic) are treated as fixed parameters in a sensitivity study.  Incorporation of these parameters into the Bayesian calibration is beyond the scope of this simple model and is left as a task for future study.

\begin{table}[h]
\begin{center}
\begin{tabular}{| c | c || c | c | c || c | c | c |}
\cline{3-8}
\multicolumn{2}{c|}{} & \multicolumn{3}{c||}{semi-diagonal} & \multicolumn{3}{|c|}{fully-covariant} \\
\hline
path-weight & $\tau_f$ & $\alpha$ & $\beta$ & \xsqrd & $\alpha$ & $\beta$ & \xsqrd \\
\hline \hline
linear & 0.1 & 0.90 & 0.27 & 0.87 & 1.80 & 0.18 & 5.93 \\
\hline
linear & 0.5 & 0.52 & 0.27 & 0.77 & 0.90 & 0.20 & 3.00 \\
\hline
linear & 0.9 & 0.50 & 0.23 & 0.74 & 0.76 & 0.18 & 2.98 \\
\hline \hline
quadratic & 0.1 & 1.30 & 0.19 & 0.74 & 2.11 & 0.11 & 2.16 \\
\hline
quadratic & 0.5 & 0.65 & 0.16 & 0.85 & 0.86 & 0.09 & 2.70 \\
\hline
quadratic & 0.9 & 0.62 & 0.12 & 0.92 & 0.75 & 0.06 & 3.17 \\
\hline  
\end{tabular}
\caption{Delta-distribution parameters and \xsqrd\ for fits to centrality-dependent $R_{AA}$} 
\label{tab:fitTrentoRAA}
\end{center}
\end{table}

\begin{figure*}[h]
  \begin{center}
    \includegraphics[width=0.9\textwidth]{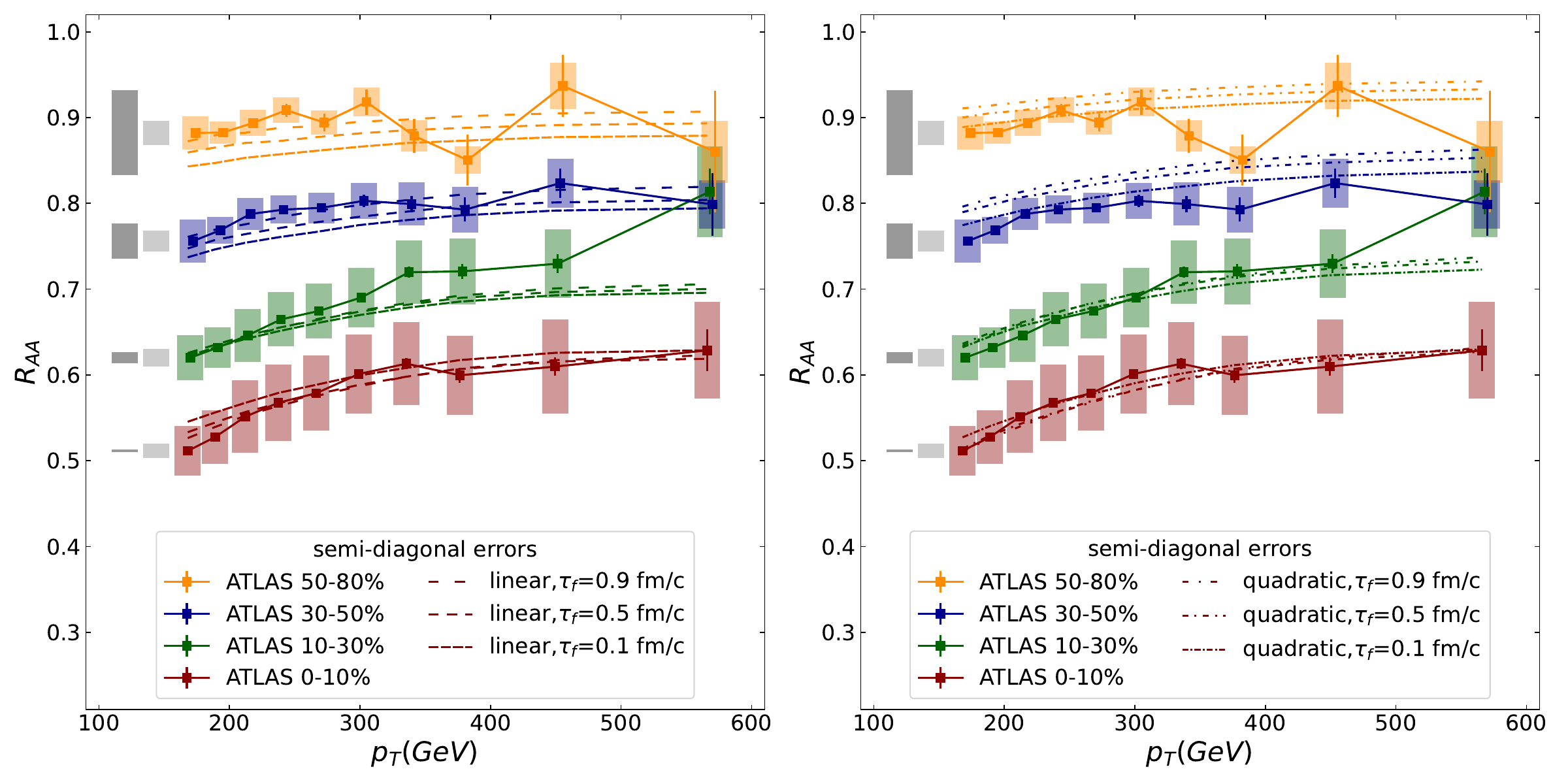}
    \caption{Simultaneous Bayesian fits to \sNN\ $R_{AA}$ measured by ATLAS for all centralities.  The left panel shows fits with a linear path-weighting and the right panel shows fits with a quadratic weighting.  Errors for nuclear-thickness (dark box) and luminosity (light grey) are fully correlated across all $p_T$ bins and centralities, but $p_T$-dependent systematic errors (colored boxes) are treated as independent.} 
    \label{fig:Trento_Ltau_dslT}
  \end{center}
\end{figure*}

A set of sensitivity studies are performed with a simultaneous Bayesian fits to the \sNN\ $R_{AA}$ measured by ATLAS for all centralities.  The simple delta-distribution $p_T$-loss is calculated by summing the energy loss over the \trento\ energy densities assuming for each of the path-length weights (linear and quadratic) illustrated in Figure~\ref{fig:TATB_heatmaps}, assuming three different values of the formation time: $\tau$=0.1,0.5,0.9 fm/c.  The calculation is performed over 650 bins in $p_T$ from 150 to 800~GeV, and each $p_T$ bin averages over 500 randomly selected di-jet paths for each of the four centrality bins.  The fit is performed using the emcee python package using 8-walkers with 10,000 samples each.  The first sensitivity study uses a covariance error matrix for which the systematic errors for the luminosity and nuclear thickness are fully correlated across all $p_T$ and centrality bins, but the $p_T$ dependent systematic errors assumed to be uncorrelated and summed in quadrature with the statistical errors.  This treatment of the systematic errors is referred to as {\em semi-diagonal}.  The results of this study are shown in shown in Figure~\ref{fig:Trento_Ltau_dslT}, with the nuclear-thickness error shown as the left-most dark grey box and the luminosity error shown as the light-grey box between the nuclear-thickness and the lowest $p_T$ bin for $R_{AA}$ value, which is also used to set the y-axis value for the grey box errors.  The $p_T$-dependent systematic errors are plotted as shaded boxes to match the color of the data points and statistical error bars.  The linear weighting is plotted in the left panel and the quadratic weighting is on the right.  The model calculations are plotted as dashed lines for the linear weighting and dash-dot lines for the quadratic weighting.  Fits with longer formation times are plotted with longer spaces between the dash and dot symbols.  This sensitivity study shows a clear preference for the longer formation time of 0.9~fm/c for a linear weights and a preference for a short formation time of 0.1~fm/c for the quadratic weights.  The shorter formation times undershoot the variation with centrality dependence for the linear case, and the longer formation times overshoot this variation for the quadratic case.  This observation is confirmed by the \xsqrd\ reported in Table~\ref{tab:fitTrentoRAA}.  The two smallest \xsqrd\ values for the optimal sensitivity combinations are equivalent, but it is also notable that all of the values are below unity, consistent with an over-fitting due to an over-estimate of the independence of the systematic errors. 

\begin{figure*}[h]
  \begin{center}
    \includegraphics[width=0.9\textwidth]{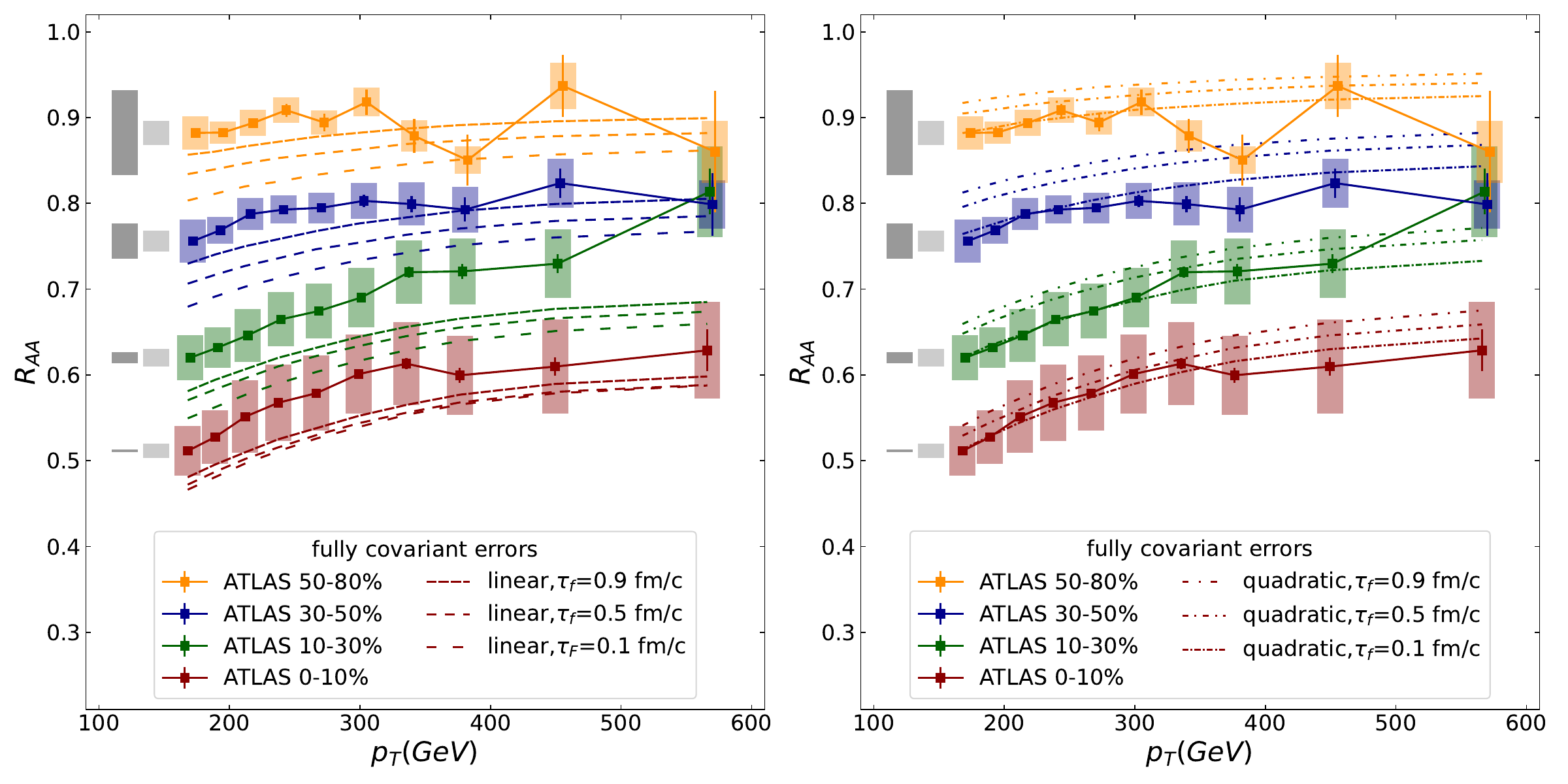}
    \caption{Simultaneous Bayesian fits to \sNN\ $\log(R_{AA})$ measured by ATLAS for all centralities.  The left panel shows fits with a linear path-weighting and the right panel shows fits with a quadratic weighting.  All contributions to systematic errors are fully correlated and then combined to construct the covariance error matrix.} 
    \label{fig:Trento_Ltau_clog}
  \end{center}
\end{figure*}

This sensitivity study was repeated using fully correlated systematic errors, for which each of the contributing components described in~\cite{10.1103/physrevc.107.054909} are assumed to be fully correlated across $p_T$ and centrality and then combined to construct a single covariant error matrix.  This treatment of the errors is referred to as {\em fully-covariant}.  The probability likelihood is calculated after a logarithmic transformation to avoid the potential for bias when fitting highly correlated errors as discussed previously and described in~\cite{10.1063/1.1945011}.
The fits, shown in Fig.~\ref{fig:Trento_Ltau_clog} and \xsqrd\ labeled as {\em fully-covariant} in in Table~\ref{tab:fitTrentoRAA} show similar trends with formation time as observed in the sensitivity study performed with partially correlated errors, except that the fits fall uniformly below the data for the linear weighting, and the corresponding \xsqrd\ values are also higher for these fits.  As will be shown in Appendix~\ref{sec:cov_impact} this may be due to a shape mis-match due to a correlated fluctuation in the data, or it may be a sign that the shape of the $R_{AA}$ curve differs from measurement when a linear weighting for the quenching is assumed.

\begin{figure}[h]
  \begin{center}
    \includegraphics[width=0.48\textwidth]{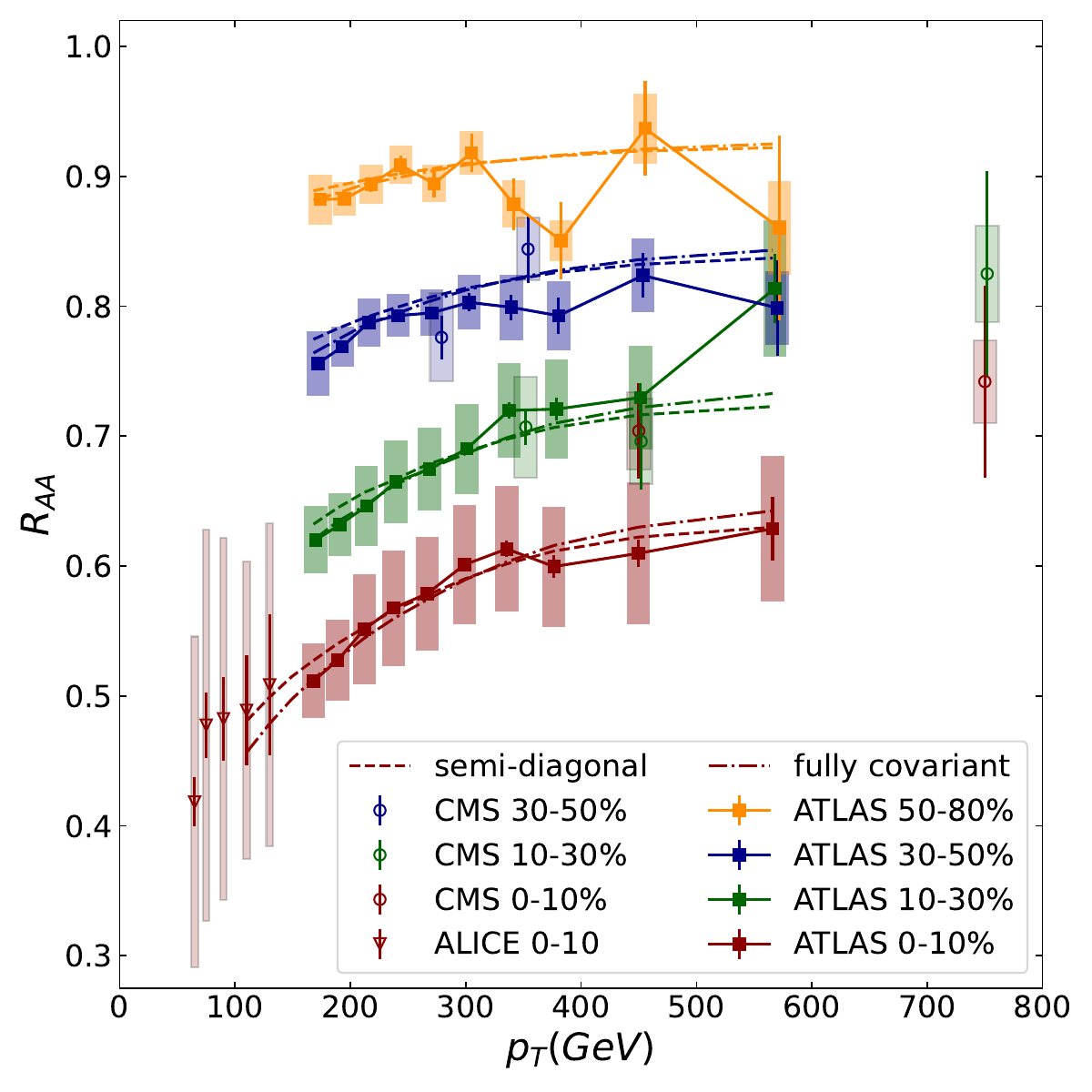}
    \caption{Bayesian fits to centrality dependent ATLAS measurements \sNN\ Pb-Pb $R_{AA}$ are compared for quadratic path weighting with $\tau=0.1$~fm/c compared to corresponding measurements by CMS and for central $R_{AA}$ measured by ALICE for charged jets.} 
    \label{fig:TrentoRAA_allexp}
  \end{center}
\end{figure}

Figure \ref{fig:TrentoRAA_allexp} shows a direct comparison between the quadratic weighting for $\tau=0.1$ between the fits to ATLAS data with semi-diagonal and fully correlated systematic errors.  Corresponding measurements for CMS~\cite{10.1007/jhep05(2021)284} and ALICE~\cite{10.1103/physrevc.101.034911} are also shown for comparison.  The ALICE jets were constructed using charged tracks for which the $p_T$ bins will have on average 2/3 of the value of similar jets measured by ATLAS and CMS, however in Figure \ref{fig:TrentoRAA_allexp} all measurements are plotted as published.  The systematic errors for ALICE include both fully correlated errors and shape errors combined in quadrature. 
This figure also includes an extrapolation of the Bayesian fits to lower $p_T$ of 100~GeV to match the lower limit of the fit to the proton-proton jet cross-section, but the higher $p_T$ of 800~GeV does not permit an extrapolation of the quenched cross-section above the highest ATLAS $p_T$ bin.  To improve clarity, the luminosity and nuclear-thickness contributions to the systematic errors are not drawn for any of the experiments.  The fits to the two different error treatments do not differ significantly within the $p_T$ range of the ATLAS measurements, but the difference become more significant for the extrapolation to lower $p_T$.

\section{Summary and Conclusions}
\label{sec:summary}

We have constructed a simple, empirical model for jet-quenching in heavy-ion collisions.  This model uses a 2-parameter product of a power-law and logarithm in the transverse momentum to describe the mean energy-loss.  The model was extended by adding an additional parameter to account for a gamma-distribution $p_T$, but this extension performed no better than a delta-function in a Bayesian fit to the \sNN\ 0-10\% central Pb-Pb $R_{AA}$ measured by ATLAS.  We conclude from this study that $R_{AA}$ measurements constrain only the mean value of $\Delta p_T$ and have little sensitivity to the shape of the $\Delta p_T$ distribution.  Extending the model to compare to the dijet asymmetry may be needed to properly constrain the gamma-distribution, as shown in ~\cite{10.48550/arxiv.2411.14552}.

The Bayesian fits were performed under several different treatments of the systematic errors.  Fits to uncorrelated errors, both statistical and combined statistical and systematic errors closely track the measured data points, but when systematic error correlations are introduced, the most-probably fit functions undershoot the measured data.  Although this behavior is consistent with a well-known phenomena in the nuclear data community known as Peele's Pertinent Puzzle~\cite{10.1063/1.1945011}, similar behavior after a log-transformation leads us to conclude that this behavior is caused by a subtle mismatch in shape between the measured data and the model.  This conclusion is supported by a series of fits to a simple functional form provided in the Appendix~\ref{sec:cov_impact}.

The simple model was further extended to describe the centrality dependence of $R_{AA}$ by incorporating collision geometry from the 2-dimensional \trento\ model using parameters determined from a Bayesian analysis with the JETSCAPE framework.  This model was used to perform a sensitivity study in which the initial \trento\ geometry undergoes boost-invariant expansion and the energy-loss is proportional to a path-integral sum over energy density assuming both a linear and quadratic weighting of the path-length through the medium using three different formation times.  We found that the range of $R_{AA}(p_T)$ for different centralities is sensitive to the choice of formation time under both of the path-length weights, with longer/shorter formation times preferred for the linear/quadratic path-weighting.  This sensitivity study was performed for two different treatments of the systematic errors.  In the {\em semi-diagonal} study, the luminosity and nuclear-thickness are fully correlated, but the systematic errors from unfolding, jet energy scale, and jet energy resolution are treated as uncorrelated.  The {\em fully covariant} study treats all contributions to the systematic errors as fully correlated across all $p_T$ and centrality bins, and these independent contributions are summed to calculate the full covariance error matrix.  In the {\em semi-diagonal} study, the data are overfit as evinced by the uniformly low values of \xsqrd\ for all parameter choices, whereas the full treatment of correlated errors produces \xsqrd\ values that are more consistent with expectations for a simple model.  In this case, the linear weighted model yields larger \xsqrd\  values and undershoots the data, indicative of a mismatch in shape.  The best fit is obtained with a quadratic weighting and a formation time of 0.1~fm/c, it is equally important to note that comparisons to the full centrality range for $R_{AA}$ may provide additional sensitivity to the study of formation times.  It would be useful to repeat this analysis with a more sophisticated model-to-data comparison in which both jet and medium formation times can be studied separately.  It is also important that future studies with physics models and frameworks incorporate error correlations to match the level of detail provided by the experiments that perform these measurements.

The python code for this model is publicly available~\cite{ez-Quench}, and can be used to generate all figures in this work.


\section*{Acknowledgments} 
This manuscript has been authored by Lawrence Livermore National Security, LLC under Contract No. DE-AC52-07NA2
7344 with the US. Department of Energy. The United States Government retains, and the publisher, by accepting the
article for publication, acknowledges that the United States Government retains a non-exclusive, paid-up, irrevocable,
world-wide license to publish or reproduce the published form of this manuscript, or allow others to do so, for United
States Government purposes.


\bibliographystyle{h-physrev5}
\bibliography{ezQ}


\begin{appendix}

\section{Covariance error matrix comparison}
\label{sec:cov_compare}

As described in Section~\ref{sec:RAAcent}, the covariance error matrix was constructed by first summing the individual contributions to the systematic error, assuming each contribution to be fully correlated across $p_T$ and centrality bins, as shown in Eq.~\ref{eq:cov_sum}.  Another common approach is to first sum the individual contributions before forming the covariance matrix according to Eq.~\ref{eq:cov_all}.  This is approach often taken when the individual contributions are not specified by the experiment.  In this case, model builders may opt for a third approach, by enforcing a correlation length across the $p_T$ fractional bin numbers, $b_i,j$ as shown in Eq.~\ref{eq:cov_part}.  This approach was taken by the JETSCAPE Collaboration in~\cite{10.1103/physrevc.104.024905}, using $l$=0.2 and $\alpha$=1.9,
\begin{eqnarray}
\Sigma_{ij}^{\rm sum} & = & \Sigma_k \sigma_i^k \sigma_j^k, 
\label{eq:cov_sum} \\
\Sigma_{ij}^{\rm all} & = & ( \Sigma_k \sigma_i^k ) ( \Sigma_k \sigma_j^k ), 
\label{eq:cov_all} \\
\Sigma_{ij}^{\rm partial} & = & ( \Sigma_k \sigma_i^k ) ( \Sigma_k \sigma_j^k ) \exp \left[ -\frac{(b_i - b_j)^\alpha}{l} \right].
\label{eq:cov_part}
\end{eqnarray}
The relative systematic error contributions for the luminosity and nuclear-thickness are both independent of $p_T$ and can be treated appropriately and separately from the assumptions implied by these equations.

\begin{figure*}[t]
  \begin{center}
    \includegraphics[width=0.75\textwidth]{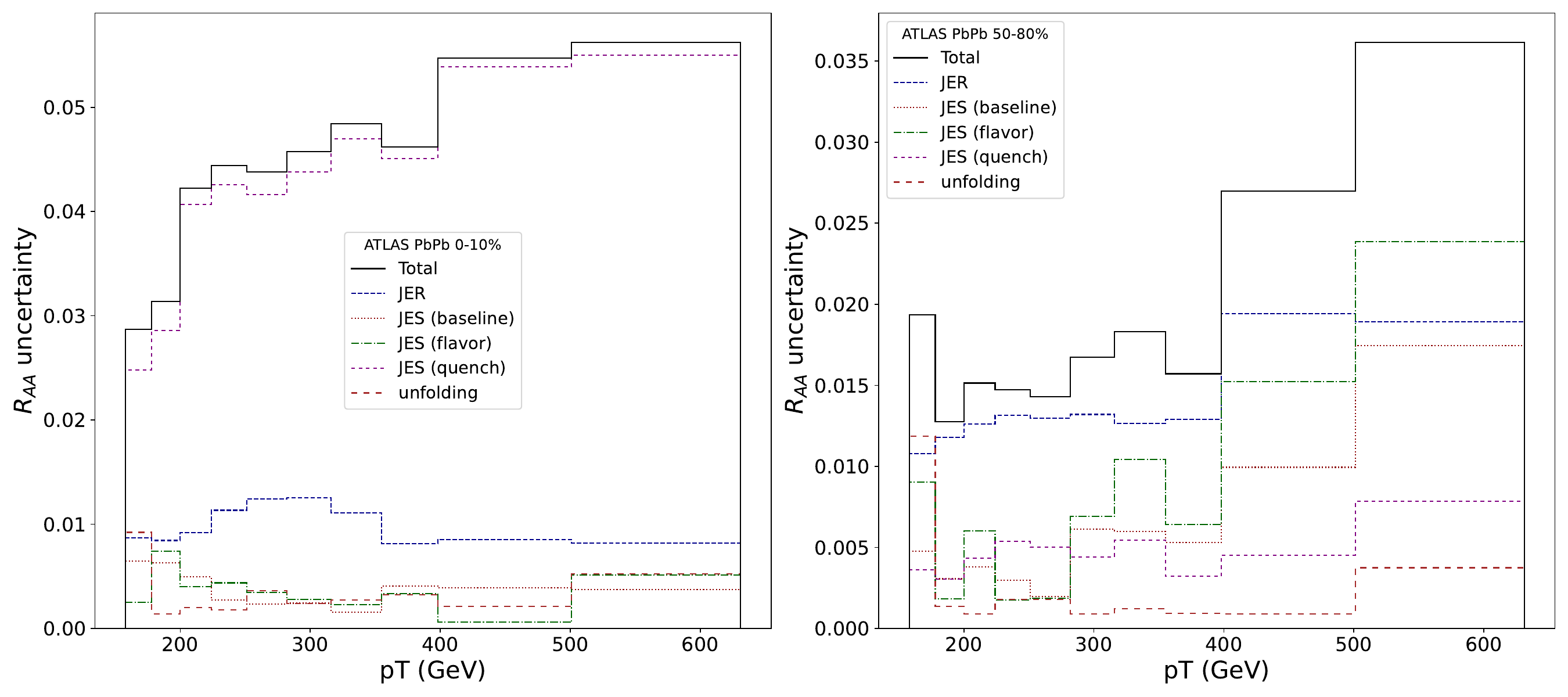}
    \caption{Systematic error contributions for ATLAS \sNN\ Pb+Pb inclusive jet $R_{AA}$ measurements for 0-10\% (left) and 50-80\% (right) centrality bins.} 
    \label{figA_ATLAS_syserr}
    \includegraphics[width=0.95\textwidth]{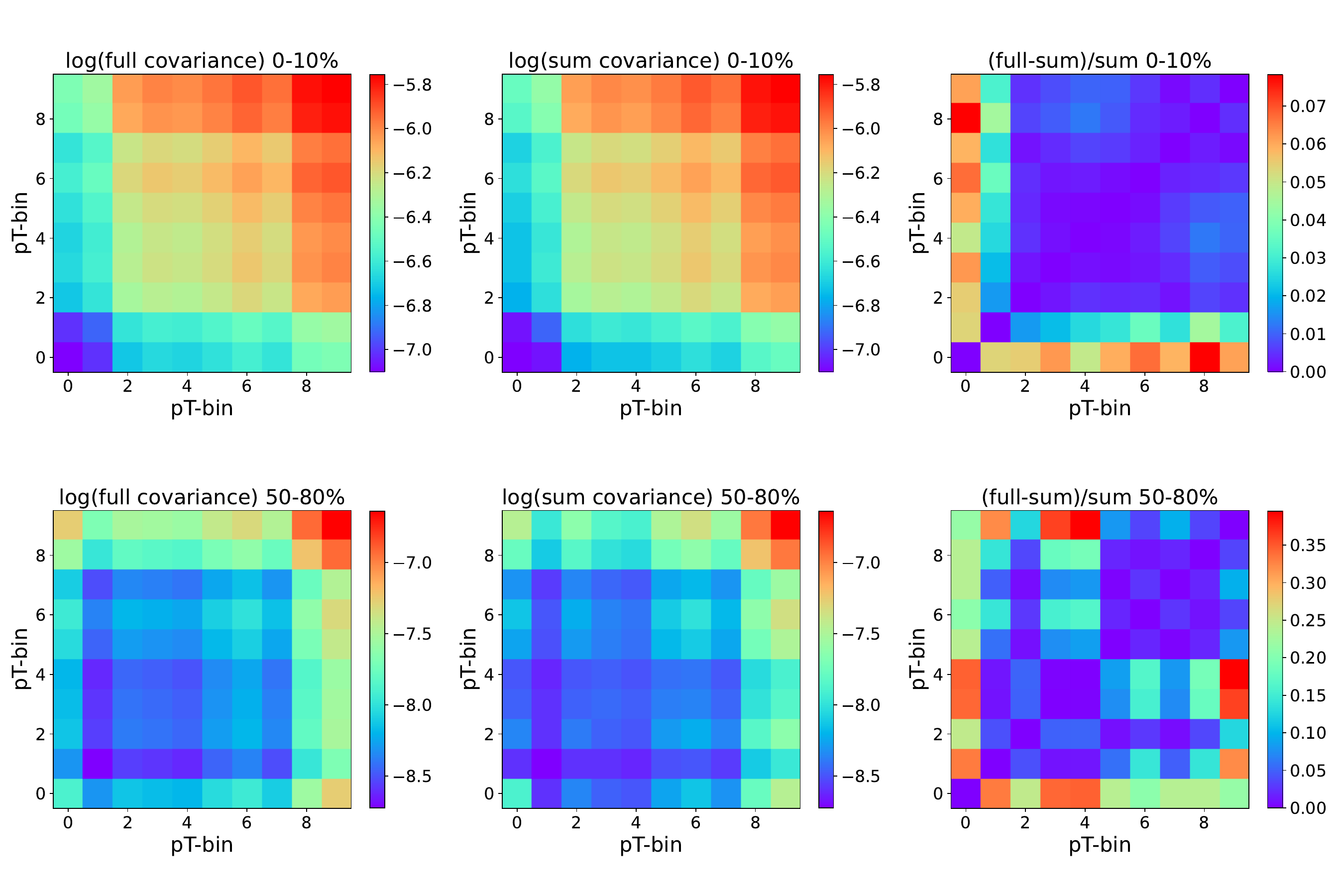}
    \caption{Covariance error matrices for ATLAS \sNN\ Pb+Pb inclusive jet $R_{AA}$ measurements for 0-10\% centrality (top) and 50-80\% centrality (bottom).  The left panels show the log of the covariance assuming full correlation in $p_T$ for all contributions, the central panels show the log covariance constructed by summing the five contributing errors sources, each assumed to be internally fully correlated, and the right panel shows the relative difference between the two methods (without taking a logarithm).}
    \label{figA_cov_compare}
  \end{center}
\end{figure*}

In this section, we compare the $\Sigma_{ij}^{\rm sum}$ and $\Sigma_{ij}^{\rm all}$ $p_T$-dependent covariance error matrices defined by Eq.~\ref{eq:cov_sum} and~\ref{eq:cov_all}, respectively.  To illustrate the source of these differences, the ATLAS systematic error components from~\cite{10.1103/physrevc.107.054909} for the 0-10\% central and 50-80\% peripheral bins are shown in Figure~\ref{figA_ATLAS_syserr}.  The Jet Energy Scale (JES) quenching component dominates the total systematic error for the central bin, but for the peripheral bin the relative contributions vary with $p_T$ and no single component dominates.  This difference between central and peripheral is shown in the covariance matrix comparison shown in Figure~\ref{figA_cov_compare}.  The differences are not easily seen when viewed side-by-side on a log scale used to capture the full range, but the relative differences plotted shown on the right panels reveal variations of order 10\% for the lowest $p_T$ bin for the 0-10\% centrality bin, and differences as high as 35\% for the peripheral bin.

\begin{figure*}[t]
  \begin{center}
    \includegraphics[width=0.48\textwidth]{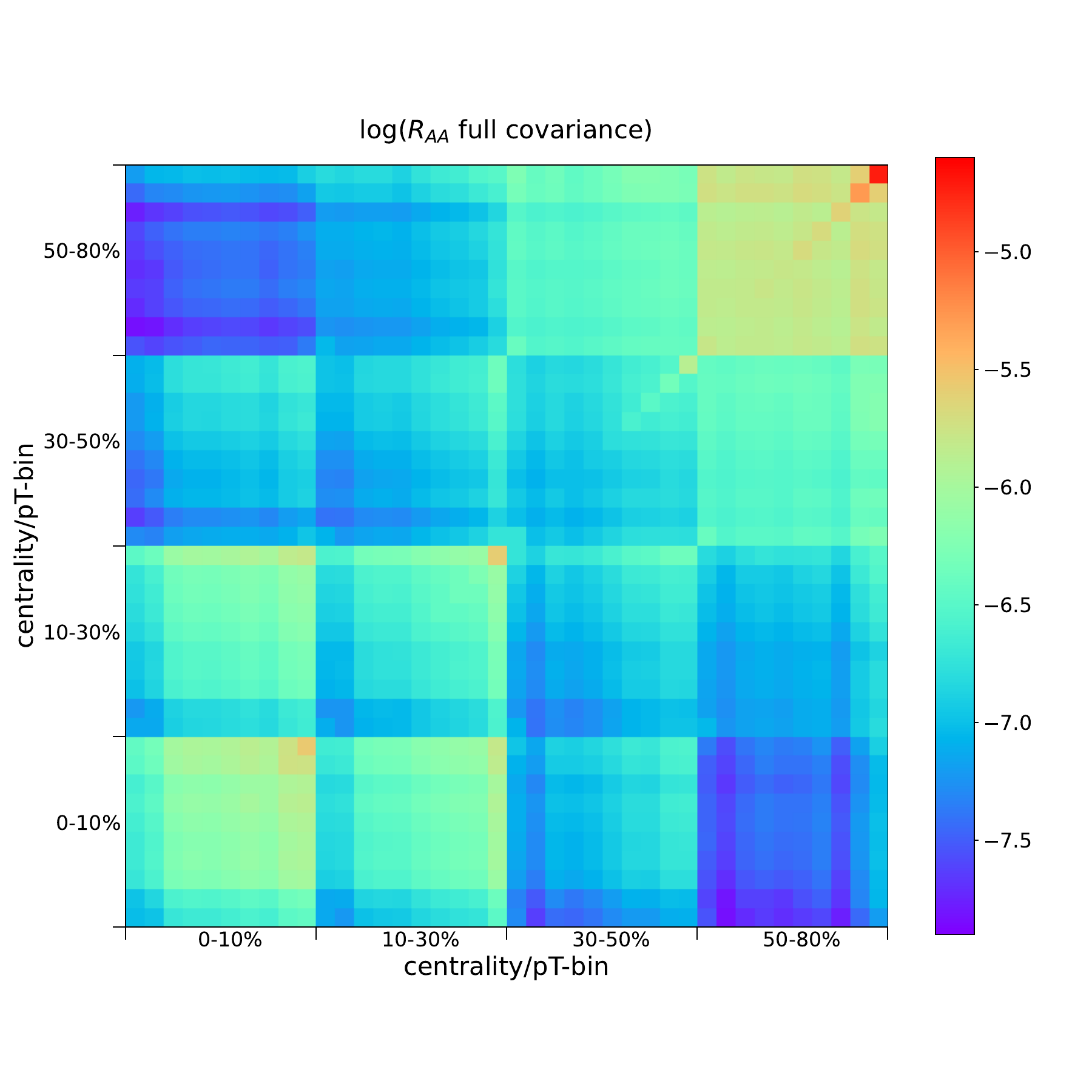}
    \includegraphics[width=0.48\textwidth]{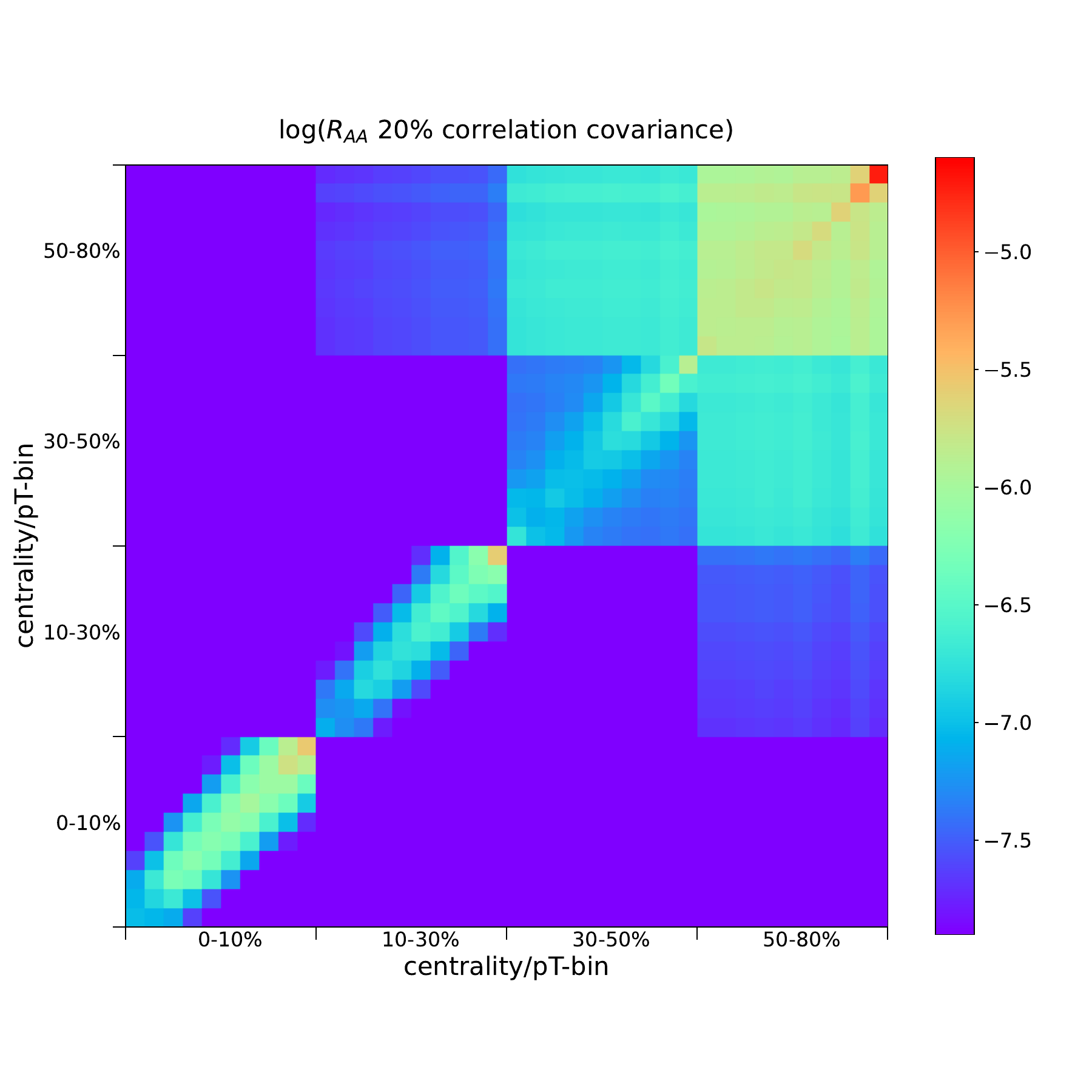}
    \caption{(Left) Full covariance error matrices for ATLAS \sNN\ Pb+Pb inclusive jet $R_{AA}$ measurements for all $p_T$ and centrality bins.
    (Right) 20\% correlation-length covariance error matrices for inclusive jet $R_{AA}$ measurements for all $p_T$ and centrality bins.}
    \label{figA_cov_complete}
  \end{center}
\end{figure*}
The log of the full covariance error-matrix used for the centrality-dependent $R_{AA}$ analysis presented in Table~\ref{tab:fitTrentoRAA} and Figures~\ref{fig:Trento_Ltau_dslT}--\ref{fig:TrentoRAA_allexp} is shown in the left panel of Figure~\ref{figA_cov_complete}.  This matrix includes all terms in $\Sigma_{ij}^{\rm sum}$ combined with the contributions from luminosity and nuclear thickness.  To complete the comparison, the right panel shows the corresponding log of the partial covariance error matrix using Eq.~\ref{eq:cov_part} for $\alpha$=2 and $l$=0.2, similar to the assumptions used by JETSCAPE~\cite{10.1103/physrevc.104.024905}.  This error matrix also includes fully correlated luminosity and nuclear-thickness errors.  The full range log-covariance extends to -9 for the off-diagonal blocks for 0-10\% and 10-30\% centrality, but the lower bound has been set to -7.9 to enable a direct comparison between the two panels.  The peripheral centrality blocks in the upper right corners appear similar (on a log scale), but the covariance values in the remaining centrality blocks are very different.  
%

\section{Correlated error impacts}
\label{sec:cov_impact}

\begin{figure}[h]
  \begin{center}
    \includegraphics[width=0.48\textwidth]{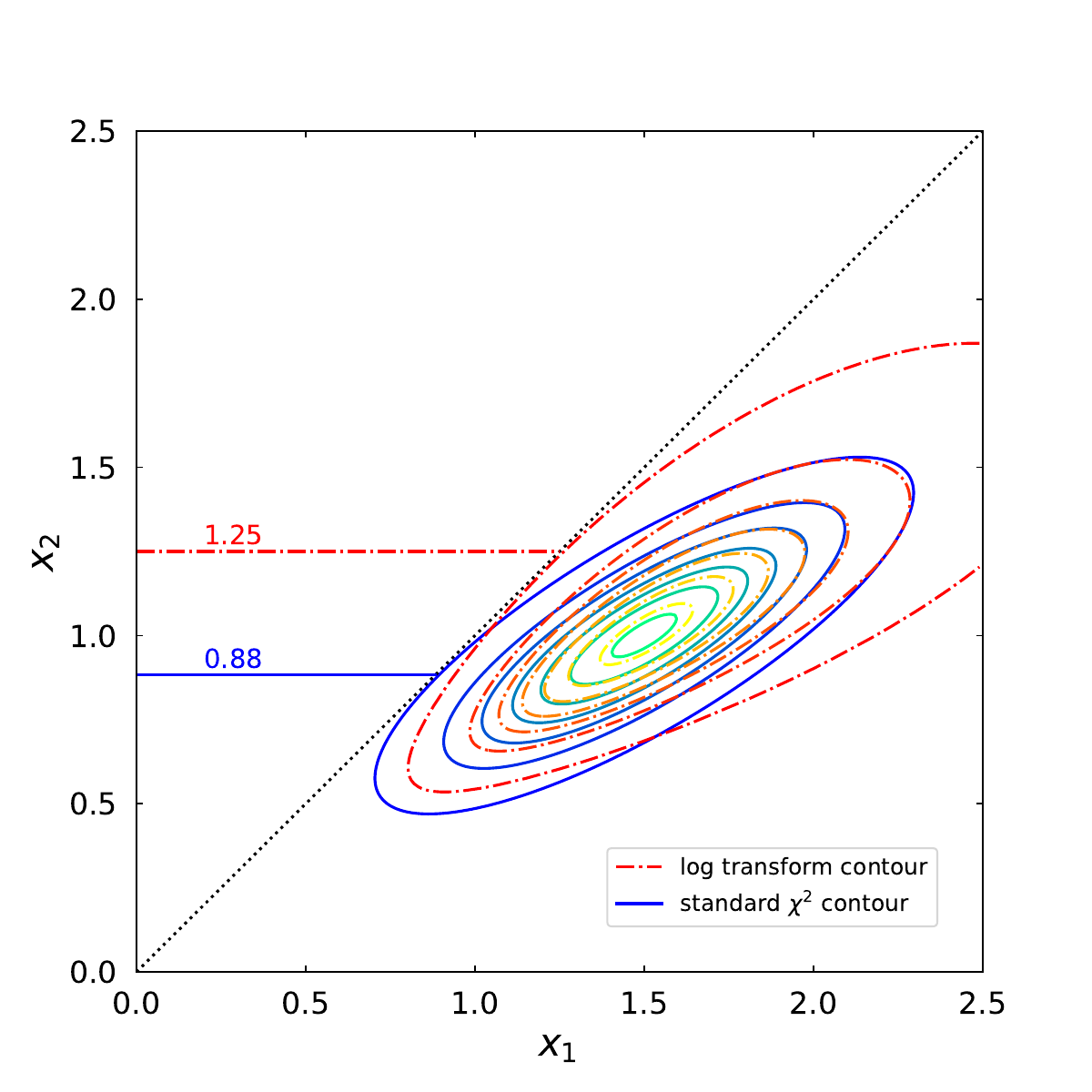}
    \caption{The joint probability distribution for measurements $m_1$ and $m_2$ (solid contours) and the joint probability distribution after a logarithmic transformation (dash-dot contours).  The requirements that the two measurements correspond to the same quantity is illustrated by the dashed-line along the diagonal.}
    \label{figB_PPP_contour}
  \end{center}
\end{figure}

\begin{figure}[h]
  \begin{center}
    \includegraphics[width=0.48\textwidth]{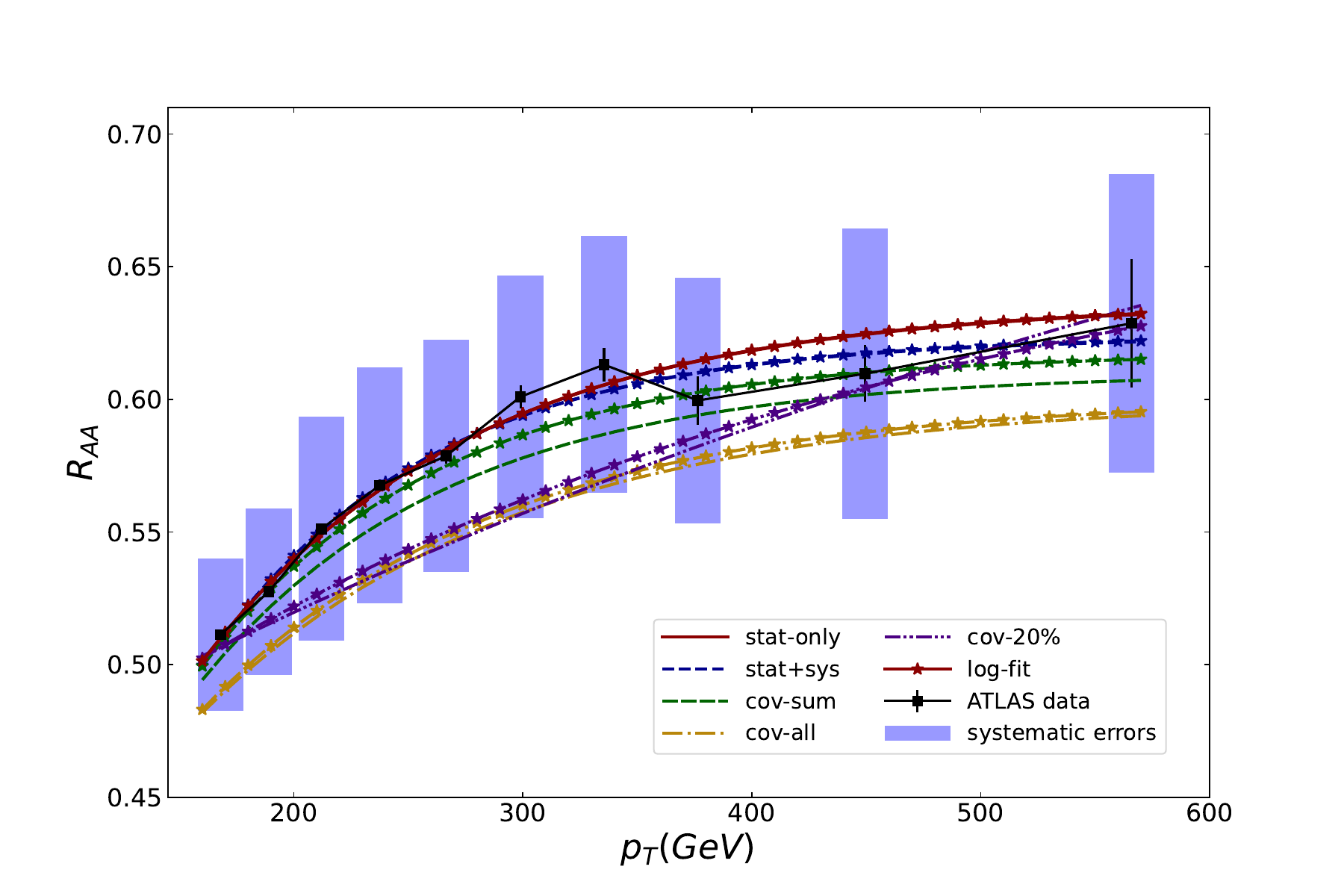}
    \caption{ATLAS jet-\RAA\ for central \sNN\ PbPb collisions fit to an exponential form for several different treatments of the errors noted in the legend.  Fits performed after taking the logarithm of both data and function are plotted in similar colors and line-styles with star symbols plotted every 10~GeV.}
    \label{figB_takeLog_compare}
  \end{center}
\end{figure}

We return to Peele's Pertinent Puzzle~\cite{10.1063/1.1945011} to study its potential impact on fits to the ATLAS data.  The puzzle is remarkably simple to describe.  Consider a quantity, $x$ for which we there are two measurements, $m_1$=1.5 and $m_2$=1.0.  Each measurement is given an independent error of 10\%, and there is a relative correlated error of 20\% between them.  It is straightforward to show that a simple $\chi^2$-minimization yields a probable-value of $x=0.882$, which lies below both of the measurements.  This counter-intuitive behavior is demonstrated graphically in Figure~\ref{figB_PPP_contour} by plotting the contours of the joint probability distribution for the two measurements.  The contours are tilted away from the diagonal line enforcing the condition that the two measurements are of the same quantity ($x_1$=$x_2$), which intersects the highest probability contour at a value of 0.882.  The authors of~\cite{10.1063/1.1945011} note that the determination whether this estimate is biased depends on the details of the measurement and provide one example, the presence of monotonically increasing background with $m_2$ preceding $m_1$, for which this estimate would not be biased.  For other cases, the authors offer several solutions to remove the potential for bias, including a suggestion to perform the $\chi^2$-minimization on the log of the model and the measurements.  Under this transformation, the relative correlated errors become additive, and the joint probability distribution, plotted as dash-dot contours in Figure~\ref{figB_PPP_contour} is less tilted near the diagonal with an intersecting value of $x$=1.25, the arithmetic average of two measurements.

To test the influence of Peele's Pertinent Puzzle on the previous analyses of the ATLAS data we fit the central jet-\RAA\ measurements to a simple exponential form, $a[1-b\exp(-p_T/c)]$ for five different treatments of the systematic errors:
\begin{enumerate}
\item {\bf stat-only:} statistical errors,
\item {\bf stat-sys:} independent combined errors,
\item {\bf cov-sum:} covariant errors (Eq.~\ref{eq:cov_sum}),
\item {\bf cov-all:} covariant errors (Eq.~\ref{eq:cov_all}),
\item {\bf cov-20\%:} covariant errors (Eq.~\ref{eq:cov_part}),
\end{enumerate}
where the {\bf cov-20} error matrix is defined with $\alpha$=2 and $l$=0.2.  All covariance error matrices include uncorrelated statistical errors, but the luminosity and nuclear-thickness contributions are neglected for this study.  The fits are performed using a standard $\chi^2$-minimization for each error matrix, and each of these fits is then repeated after taking the logarithm of the data and fit function with results for all fits shown in Figure~\ref{figB_takeLog_compare}.  The different error treatments are plotted in different line styles as noted in the legend, and the log fits use the same line-styles and colors but with star symbols placed every 10~GeV.  The fits to independent statistical and combined errors are unchanged by the log transformation.  Although the log transformation fit is closer to the data for the summed covariance error matrix, the differences are negligible for the other treatments of the correlated errors.  From this comparison we conclude that lower values of the fit function may be due more to shape differences in the $p_T$ dependence between the data with correlated errors and the exponential function.

\begin{figure}[h]
  \begin{center}
    \includegraphics[width=0.48\textwidth]{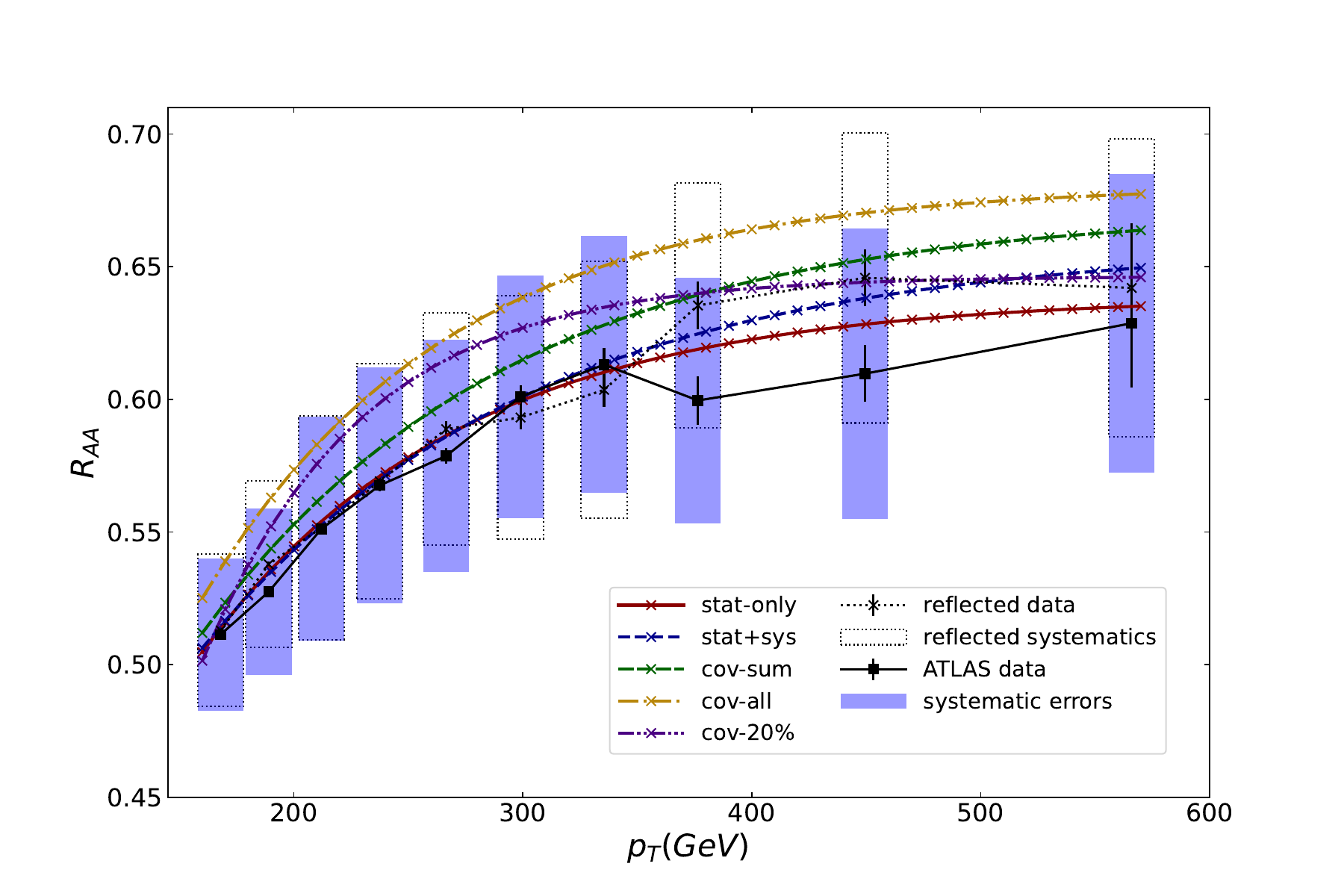}
    \caption{ATLAS jet-\RAA\ for central \sNN\ PbPb collisions (squares) and the same data reflected about the previous exponential fit using only statistical errors (x).  The reflected data are re-fit to the same exponential form for each treatment of the systematic errors noted in the legend.}
    \label{figB_flipData_compare}
  \end{center}
\end{figure}

To further test this conclusion we perform one more series of fits using the same fit function and covariance error matrices, but reflecting (or flipping) the data about the previous fit to the data with only statistical errors, as  shown in Figure~\ref{figB_flipData_compare}.  The new fits with independent errors follow the new data, but all fits using correlated errors are now mostly above the new reflected data, confirming the previous assertion that in this specific example, the behavior of these fits is sensitive subtle shapes in the data and their errors and not exclusively a manifestation of Peele's Pertinent Puzzle.

\end{appendix}

\end{document}